\documentclass[pre,twocolumn,fleqn,showpacs,superscriptaddress,showkeys]{revtex4}
\usepackage{graphicx}
\usepackage{amssymb}
\usepackage{amsmath}
\usepackage{bm}
\usepackage{color}
\begin{document}

\newcommand{\BEQ}{\begin{equation}}
\newcommand{\EEQ}{\end{equation}}
\newcommand{\BEA}{\begin{eqnarray}}
\newcommand{\EEA}{\end{eqnarray}}
\newcommand{\BEF}{\begin{figure}}
\newcommand{\EEF}{\end{figure}}

\setcounter{page}{0}
\title[]{Condensation and Clustering in the Driven Pair Exclusion Process}
\author{Sang-Woo \surname{Kim}}
\affiliation{Department of Physics, University of Seoul, Seoul 130-743,
Korea}
\author{Jae Dong \surname{Noh}}
\affiliation{Department of Physics, University of Seoul, Seoul 130-743,
Korea}
\affiliation{School of Physics, Korea Institute for Advanced Study, 
Seoul 130-722, Korea}
\email{jdnoh@uos.ac.kr}

\date{\today}

\begin{abstract}
We investigate particle condensation in a driven pair exclusion process on
one- and two- dimensional lattices under the periodic boundary condition.
The model describes a biased hopping of particles subject to a pair
exclusion constraint that each particle cannot stay at a same site with 
its pre-assigned partner. The pair exclusion causes a mesoscopic
condensation characterized by the scaling of the condensate size 
$m_{\rm con}\sim N^\beta$ and the number of condensates 
$N_{\rm con}\sim N^\alpha$ with the total number of sites $N$. 
Those condensates are distributed randomly without hopping bias. 
We find that the hopping bias generates a spatial correlation among
condensates so that a cluster of condensates appears.
Especially, the cluster has an anisotropic shape 
in the two-dimensional system. 
The mesoscopic condensation and the clustering are studied 
by means of numerical simulations.  
\end{abstract}



\maketitle

\section{Introduction}\label{sec:1}
In a driven diffusive system, hopping bias in dynamics drives a system out
of equilibrium and can lead to a variety of effects.
The hopping bias is irrelevant for noninteracting or 
single-particle systems. Biased diffusion of a particle, for example,
is equivalent to unbiased diffusion in a co-moving frame. Particle
interaction can make a difference. 
Consider, as an example, the asymmetric simple exclusion
process~(ASEP)~\cite{Derrida98} in
one dimension~(1D), in which particles hop under the constraint that each
site can be occupied by at most a single particle~[exclusion interaction].
When particles hop symmetrically, the system belongs to the
Edwards-Wilkinson universality class in which the dynamic correlation length 
$\xi$ grows in time $t$ as $\xi \sim t^{2}$~\cite{Edwards82}. 
With a hopping bias, however,
the system exhibits a scaling $\xi \sim t^{3/2}$ of the
Kardar-Parisi-Zhang universality class~\cite{Kard86}. 

In general, the hopping bias can also change the nature of a stationary state
probability distribution. We focus on systems displaying a condensation
transition. Condensation occurs in various forms such as Bose-Einstein 
condensation, traffic jams, hub formation in evolving networks, and so
on~\cite{Evans05}. 
Condensation can be studied in the context of driven diffusive
systems~\cite{Evans05,Majumdar98}, and the effect of hopping bias has 
been studied in those systems~\cite{Rajesh01,Rajesh02,Godreche03,Luck07}.

The zero range process~(ZRP) is useful in studying
particle condensation~\cite{Spitzer70,Jeon00,Evans05}. 
Consider, for simplicity, a one-dimensional ring of $N$ sites
$i=1,2,\cdots,N$ under a periodic boundary condition. 
There are $M$ particles, and the occupation number at site $i$ is denoted as
$m_i=0,1,2,\cdots$. A particle at site $i$ selected randomly 
decreases by one~($m_i \to m_i-1$) at a rate $u(m_i)$ and then hops to 
a neighboring site $j=i+1$ with probability $p$ or $j=i-1$
with probability $q=1-p$. The jumping rate depends on the occupation number 
at a source site, whose functional form reflects the nature of on-site 
interactions among particles. For example, if particles hop 
independently, the jumping rate should be linearly 
proportional to the occupation number, $u(m) \propto m$. 
A jumping rate function $u(m)$
growing sublinearly or decreasing in $m$ corresponds to an 
attractive interaction. 

The stationary state of the ZRP is known exactly. The probability
$P_{st}(\bm{m})$ to find the system in a configuration 
$\bm{m}=(m_1,\cdots,m_L)$ in the stationary state is given by a product 
form
\BEQ\label{ZRP_Pst}
P(\bm{m}) = \frac{\delta(M-\sum_{i=1}^N m_i)}{Z(N,M)} \prod_{i=1}^N \left[
\frac{1}{\prod_{l=1}^{m_i} u(l)}\right]
\EEQ   
with a normalization factor $Z(N,M)$.
The exact solution allows
one to understand the condition for condensation. Suppose that the jumping
rate is given by $u(m)=1+b/m$. When $b<2$, particles are distributed 
uniformly at any value of particle density $\rho=M/N$. On the other hand, 
when $b>2$, there emerges a single macroscopic
condensate of size $m_{\rm con} = (\rho-\rho_c)N$ for $\rho>\rho_c$ with
$\rho_c=1/(b-2)$. The ZRP has the same stationary state 
at all values of $p$ and $q$ irrespective of the hopping bias.

An interesting generalization of the ZRP was considered in Ref.~\cite{Luck07},
where the particle jumping rate is constant, but a jump is accepted with 
probability $v(m)$ depending on the occupation number
$m$ at target sites.
Such a model is called the target process~(TP). Without a hopping 
bias~($p=q$), the stationary state of the TP is mapped to that of the ZRP
with jumping rate $u(m+1) = 1/v(m)$. 
Hence, the unbiased TP displays macroscopic condensation under an 
appropriate condition, e.g., 
$v(m) = 1/(1+b/(m+1))$ with $b>2$ and $\rho>1/(b-2)$.
However, when a hopping bias with $p\neq q$ exists, the mapping breaks
down, and the stationary state of the TP is not given by the product form.
Moreover, the hopping bias destroys the condensation~\cite{Luck07}.

The conserved mass aggregation~(CMA) model also exhibits a 
condensation phenomenon~\cite{Majumdar98}. 
In this model, all particles at a site $i$
hop~($m_i\to 0$) to a neighboring site at the unit rate or a single particle
is chipped off~($m_i\to m_i-1$) to a neighboring site at a constant 
rate $\omega$. 
Particles at different sites aggregate through the former process while
they are scattered out by the latter. Competition between them 
results in condensation when $\rho\ge \rho_c(\omega) = \sqrt{\omega+1}-1$ 
in all dimensions when the hopping and the chipping are 
symmetric~\cite{Rajesh01}. However, a bias in
the hopping and the chipping was shown to inhibit condensation in
one-dimension~\cite{Rajesh02}.

In this work, we study a particle condensation phenomenon in a driven pair
exclusion process~(PEP). The PEP was first introduced in Ref.~\cite{Kim10}
as a model for hub formation in evolving networks~\cite{Kim08,Kim09,Kim10a}.
In the PEP, particles hop under the so-called pair exclusion constraint,
which will be explained later. With symmetric hopping, the system in the
stationary state exhibits an intriguing condensation state 
characterized by multiple {\em mesoscopic} condensates 
of size $m_{\rm con} \sim N^{1/2}$ and of number $N_{\rm con}\sim N^{1/2}$, 
where $N$ is the total number of particles with a logarithmic correction~\cite{Kim10}. 
Those condensates are distributed randomly without any spatial correlation. 
We will investigate the effect of the hopping bias on the nature of 
mesoscopic condensation.

The paper is organized as follows: In Sec.~\ref{sec:2} we introduce the
PEP with and without a hopping bias. The condensation phase transition 
of the driven PEP has been investigated via numerical simulations, 
results of which are presented in Sec.~\ref{sec:3}. We summarize the paper
in Sec.~\ref{sec:4}.

\section{Driven Pair Exclusion Process}\label{sec:2}
There are $M$ particles on a $d$-dimensional hypercubic lattice of $N=L^d$ 
sites under a periodic boundary condition. 
The occupation number at site $i$ is denoted as $m_i$. 
Let us assume that 
there are $M/2$ distinct particle species and that each species 
has two elements. Pair exclusion means that no pair of particles 
of the same species is not allowed to stay at a same site. Such an
interaction appears naturally in evolving networks~\cite{Kim10a}.

\begin{figure}
\includegraphics[width=\columnwidth]{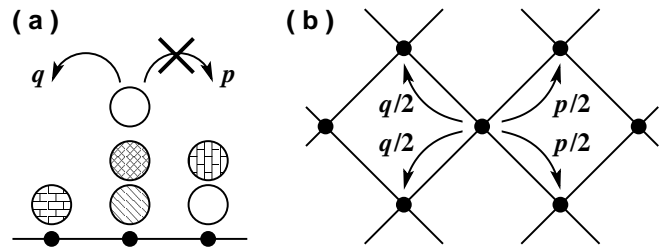}
\caption{The driven pair exclusion process in one (a) 
and two dimensions (b). Particle species are distinguished with 
filled patterns in (a). Pair exclusion forbids the particle 
represented by an empty circle to hop to the right.}
\label{fig1}
\end{figure}

The particle hopping dynamics is given as follows:
At each time step,
(i) we select a source site $i$ at random from among $N$ sites. 
(ii) When $m_i>0$, we select one particles from among $m_i$ particles and 
attempt a particle hopping with the probability 
\BEQ \label{u_m}
u(m_i) = \left(1+\frac{b}{m_i}\right) / (1+b) . 
\EEQ 
(iii) There are $2d$ possible target sites 
$\{i\pm \hat{e}_k | k=1,\cdots,d\}$, where $\hat{e}_k$ denotes
the unit vector in the $k$th direction. 
The target site is chosen from among {\em forward} sites 
$\{i+\hat{e}_k|k=1,\cdots,d\}$ with
probability $p$ or from among {\em backward} sites 
$\{i-\hat{e}_k|k=1,\cdots,d\}$ with probability $q=1-p$. 
(iv) The hopping attempt is accepted only if it does not violate pair
exclusion. The hopping is symmetric when $p=q=1/2$ while it is biased to 
the forward~(backward) direction when $p>q$~($p<q$). The particle dynamics
is illustrated in Fig.~\ref{fig1}.

Due to pair exclusion, a hopping attempt is accepted or rejected
depending on a particle's species distribution. 
In Ref.~\cite{Kim10}, the accepting probability
was shown to be approximately given by
\BEQ \label{P_target}
v(m) \simeq 1 - \frac{m}{M} \ ,
\EEQ
where $m\ll M$ is the occupation number at a target site. 
Using the approximation, we can map the PEP to a driven diffusive system 
whose particle hopping probability from site $i$ to $j$ is given by
\BEQ \label{Wvu}
W_{ji} = p_h v(m_j) u(m_i)
\EEQ
with the hopping probability $p_h=p~(q)$ to the forward~(backward) 
direction.

When the hopping is symmetric~($p=q$), the
model is solvable, and its stationary state probability distribution 
is given by a product form as in Eq.~(\ref{ZRP_Pst})~\cite{Kim10}. 
In fact, the model has the same stationary
state as the ZRP with the hopping rate function given by
\BEQ\label{u_zrp}
u_{ZRP}(m) = u(m) / v(m-1) \propto 1 + \frac{b}{m} + \frac{m}{M} \ .
\EEQ
Note that the additional factor $\frac{m}{M}$ accounts for
pair exclusion. Such a factor is irrelevant when
$m=\mathcal{O}(1)$, but it plays a crucial role in the condensation
phase. It suppresses a macroscopic condensate, and 
multiple mesoscopic condensates of size $m_{\rm con} \sim N^{1/2}$ and of number
$N_{\rm con} \sim N^{1/2}$ appear~\cite{Angel05,Angel07,Schwarzkopf08,Kim10}. 
Those mesoscopic condensates are spatially uncorrelated because
the probability distribution has a factorized product form.

In this work, we investigate the driven PEP with $p=1$ and $q=0$. The
hopping bias invalidates the mapping of the PEP to the ZRP; hence, 
the stationary state probability distribution is not given by a product
form.
One obvious question is whether macroscopic or mesoscopic condensation 
occurs or not. Another interesting question is about a spatial
correlation. Since the stationary state is not given by a factorized form, 
a spatial correlation in the particle distribution exists.
Numerical simulation results on these issues are presented in the following
section.

\section{Numerical Results}\label{sec:3}
We have performed extensive Monte Carlo simulations for the driven PEP in
$d=1$ and $d=2$ dimensions. Particles are allowed to hop only in the preferred
forward direction~($p=1$ and $q=0$) (see Fig.~\ref{fig1}). 
We adopt the jumping probability in Eq.~(\ref{u_m}) with $b=4$. 
This particular value of $b$ is
chosen because it allows mesoscopic condensation for the symmetric 
PEP~\cite{Kim10}. We start with $M$ particles
distributed randomly on $N=L^d$ sites, and data are measured in
the stationary state over a time interval $T\geq 10^8$. 
The system sizes are up to $L=16000$ in 1D and $L=100$ in 2D.
The condensation transition is examined with the occupation number
distribution
\BEQ
P(m) = \frac{1}{L^d} \left\langle \sum_{i=1}^{L^d} \delta(m_i,m) 
\right\rangle \ ,
\EEQ
which is the probability of a site having $m$ particles.

\BEF[th]
\includegraphics*[width=\columnwidth]{fig2.eps}
\caption{The occupation number distribution $P(m)$ is shown 
at various system sizes $N=L^d$ for 1D in (a) and 2D in (b). 
Parameter values are $\rho=4$ and $b=4$. 
The inset shows $P(m)$ for systems of size $N=16000$ in 1D and $N=100^2$
in 2D when the particle density $\rho=1/4$, $1/2$, and $4$,
which are below, equal to, and above the critical density
$\rho_c=1/2$, respectively. The symbols represent the critical distribution
of the corresponding ZRP given in Eq.~(\ref{P_c})  with $b=4$.} \label{fig2}
\EEF

When the particle density is low, the distribution function decays 
exponentially~(see the insets in Fig.~\ref{fig2}) in $m$. 
Without pair exclusion, the PEP reduces to the ZRP, which does not 
show condensation at low particle density. 
Since pair exclusion suppresses condensation, naturally,
condensation does not occur.

When the particle density is so low that there is no condensate in the system, 
the effect of pair exclusion represented by Eq.~(\ref{P_target}) 
can be negligible in the infinite size limit. Namely, the PEP 
becomes equivalent to the ZRP in the normal phase without condensates.
We expects that the equivalence persists up to the critical density
$\rho_c=1/(b-2)$ at which the ZRP undergoes a condensation
transition~\cite{Evans05}.
In the insets of Fig.~(\ref{fig2}), we compare the occupation number
distribution at $\rho=\rho_c$ with the critical occupation number
distribution of the ZRP~\cite{Evans05}
\BEQ\label{P_c}
P_c(m) \propto \frac{\Gamma(m+1)}{\Gamma(m+b+1)} 
\EEQ
with $\Gamma(x)=(x-1)!$, which scales as $P_c(m)\sim m^{-b}$ for large $m$.
The numerical data are in good agreement with the critical distribution with
$b=4$. This comparison shows that the driven PEP undergoes a condensation
transition at the same threshold $\rho_c = 1/(b-2)$.

\BEF
\includegraphics*[width=\columnwidth]{fig3.eps}
\caption{Power-law scaling of the size of a typical condensate
$m_{\rm con}$~($\square$), the number of condensates 
$N_{\rm con}$~($\diamond$), and the total number of particles 
belonging to the condensates $m_{\rm total}$~($\circ$) in 1D (filled symbols) 
and 2D (empty symbols). Parameter
values are $b=4$ and $\rho=4$. Solid line are guides for the eye.}\label{fig3}
\EEF

When the density is high, a broad peak in $P(m)$ appears.
The peak position moves to the right as $L$
increases, as shown in Fig.~\ref{fig2}. The peak represents condensates. 
The peak is broad, so we quantify the size of a typical condensate,
$m_{\rm con}$, with the highest peak position. 
The dependence of the condensate size on the number of lattice sites 
is plotted in Fig.~\ref{fig3}. It follows a
power-law scaling as
\BEQ\label{beta}
m_{\rm con}(N) \sim N^{\beta} 
\EEQ
with $\beta \simeq 0.56$ in 1D and $\beta \simeq 0.62$ in
2D. The condensate is not macroscopic but mesoscopic with $0<\beta<1$.

We note that the spectral weight of the condensate peak is much larger 
than $1/N$, which implies that there are multiple condensates. 
We quantify the number of condensates, $N_{\rm con}$, 
from the total spectral weight of the peak. 
There is a local minimum in $P(m)$ separating the occupation number
into two regions. We estimated $N_{\rm con}$ as the total spectral weight 
beyond the local minimum multiplied by $N$. Numerical data 
in Fig.~\ref{fig3} show that it also follows a power-law scaling as
\BEQ\label{alpha}
N_{\rm con} \sim N^\alpha 
\EEQ
with $\alpha \simeq 0.51$ in 1D and $\alpha \simeq 0.34$ in 2D.
The total number of particles belonging to the condensates is proportional
to $N$.
 
We have shown that the driven PEP exhibits mesoscopic condensation
characterized by the scaling $m_{\rm con}\sim N^\beta$ and 
$N_{\rm con} \sim N^\alpha$.  
It contrasts with the CMA and the TP in which 
condensation does not occur in the presence of a hopping bias.
Although the driven PEP exhibits a similar type of mesoscopic condensation
as the symmetric PEP, significant differences exist.
The symmetric PEP can be approximated as a driven diffusive system with 
the hopping probability given in Eq.~(\ref{Wvu}), which can be mapped to 
the ZRP. Since the stationary state of the ZRP is given by a factorized 
product form, no spatial correlation exists. 
Consequently, the ZRP in all dimensions has the same property~\cite{Kim10}. 
The variation of $\alpha$ and $\beta$ with respect to $d$ in the driven PEP
suggests that a spatial correlation does matter.

\BEF
\includegraphics*[width=\columnwidth]{fig4.eps}
\caption{(a) Occupation number distribution function $C(r)$ around 
the largest condensate in 1D with $b=4$ and $\rho=4$. 
The inset shows a snapshot of an occupation number distribution. 
(b) Scaling plot of $C(r)/N^\beta$ against
$r/N^\delta$ with $\beta=0.56$ and $\delta=0.30$.}\label{fig4}
\EEF

The spatial correlation is clearly seen from the spatial distribution of the 
occupation number. 
When condensation occurs, one can locate the site $i_M$ 
at which the occupation number is maximum. Then, one can measure the mean
occupation number $C(r)$ at site $j=i_M+r$ displaced from $i_M$ by $r$.
The distribution function is plotted for 1D at $b=4$ and $\rho=4$ in
Fig.~\ref{fig4} (a). The occupation number is high near $r=0$ and then
decays to a constant value at large $r$. This shows that 
mesoscopic condensates are bound to each other to form a cluster. We
speculate that the clustering may originate from the jamming of driven
condensates due to pair exclusion.

The occupation number distribution turns out to follow a scaling form
\BEQ\label{C1d}
C(r) = N^{\beta} \mathcal{F} (r/N^{\delta}) 
\EEQ 
with $\beta=0.56$ and $\delta=0.30$~(see Fig.~\ref{fig4}(b)). The scaling
function decays exponentially.
There are $N_{\rm con}
\sim N^\alpha$ condensates, so one may expect that $\delta = \alpha$. 
However, the numerical analysis in Fig.~\ref{fig4}(b) yields a value of
$\delta=0.30$, which is smaller than $\alpha=0.51$. The discrepancy may be
explained by the assumption that there are $N_{\rm cl} \sim N^{\alpha-\delta}$ 
such condensate clusters.
In fact, the inset of Fig.~\ref{fig4}(a) shows that there are a few 
condensate clusters. With small values of $(\alpha-\delta)\simeq 0.2$, 
the expected number of clusters is small, and we could not verify 
the power-law scaling numerically.

\BEF[th]
\includegraphics*[width=\columnwidth]{fig5.eps}
\caption{Scaling of the occupation number along the directions (a) parallel 
and (b) perpendicular to the bias in 2D with $b=4$ and $\rho=4$.}
\label{fig5}
\EEF

We also study the clustering of mesoscopic condensates in 2D.
Measuring the occupation number distribution $C(r)$ at all lattice sites
$j = i_M + r$ with $r = x \hat{e}_1 + y \hat{e}_2$,
we found that all condensates form a single cluster.
Interestingly, the condensate cluster has an elongated shape. 
Presented in Fig.~\ref{fig5} are numerical
data $C_{\parallel}(x)$, the occupation number at sites
$j=i_M + x(\hat{e}_1+\hat{e}_2)$ along the direction parallel to
the bias, and $C_\perp(x)$, the occupation number at sites $j=i_M+x(\hat{e}_1 -
\hat{e}_2)$ along the direction perpendicular to the bias. They are fitted
well to the scaling form
\BEQ\label{C2d}
C_{\parallel,\perp}(x) = N^{\beta} \mathcal{G}_{\parallel,\perp} (
x/N^{\delta_{\parallel,\perp}}) \ .
\EEQ
The exponent $\beta=0.62$ is the same as the one for the
condensate size. On the other hand, the two exponents describing the
characteristic width of the cluster have different values of $\delta_\parallel
\simeq 0.25$ and $\delta_\perp \simeq 0.1$. The total number of mesoscopic
condensates inside the cluster scales as $N^{\delta_\parallel+\delta_\perp}
\sim N^{0.35}$, which is close to the previous estimate $N^{\alpha}$ with
$\alpha = 0.34$.

We add a few remarks on the shape of the occupation number distribution
function. We observe that the distribution is symmetric as 
$C(r)=C(-r)$ in 1D. In 2D, the distribution in the transverse direction
is trivially symmetric as $C_\perp(x) = C_\perp(-x)$. 
However, it is asymmetric in the longitudinal direction as 
$C_\parallel (x) \neq C_\parallel (-x)$. The origin of the asymmetry is not
understood yet.

From the spatial distribution of the occupation number, we conclude that
there is a clustering of condensates in the driven PEP.
Condensate clustering was observed in interacting particle
systems~\cite{Evans06,Waclaw09}, where the clustering was caused by a
particle attraction. 
Clustering in the driven PEP has a different origin.
Clustering does not occur in the unbiased PEP where mesoscopic condensates 
are distributed randomly. 
When there is a hopping bias, there is congestion due to the
pair exclusion. The jamming leads to  clustering.

\section{Summary}\label{sec:4}
We have investigated particle condensation in the driven PEP 
in 1D and 2D. The PEP with symmetric hopping displays mesoscopic
condensation, which is characterized by uncorrelated condensates of number $N_{\rm
con} \sim N^{\alpha}$ and of size $m_{\rm con}\sim N^{\beta}$ with
$\alpha=\beta=1/2$ in all dimensions~\cite{Kim10}. 
The driven PEP also exhibits mesoscopic condensation. 
However, a fundamental difference exists.  
The driven PEP is characterized by a spatial correlation that is manifested
in the $d$-dependent exponents and the clustering of condensates. The clustering is
a consequence of jamming caused by the combined effect of the hopping bias and
the pair exclusion. 
The shape of the condensate cluster follows the scaling form of Eqs.~(\ref{C1d}) and
(\ref{C2d}). It is interesting to note that the cluster is anisotropic in 2D.
Our results show that the spatial correlation leads to rich behaviors of 
condensation phenomena, which need to be investigated thoroughly in the
future.

\begin{acknowledgments}
This work was supported by the University of Seoul 2009 Research Fund.
\end{acknowledgments}

\end{document}